\documentclass{jpsj2}
\usepackage{graphicx}

\title{Inhomogeneous Electronic Distribution in High-T$_{\mathrm{c}}$ Cuprates}
\author{Shigeru Koikegami$^1$
\footnote{E-mail: shigeami@secondlab.co.jp}, Masaru Kato$^2$, and Takashi Yanagisawa$^3$}
\inst{$^1$Second Lab, LLC, Tsukuba, Ibaraki 305-0045, Japan\\
$^2$Department of Mathematical Sciences, 
Osaka Prefecture University, Sakai 599-8531, Japan\\
$^3$Electronics and Photonics Research Institute, 
AIST Tsukuba Central 2, Tsukuba, Ibaraki 305-8568, Japan}

\abst{We theoretically investigate the doping evolution of the electronic state of high-T$_{\mathrm{c}}$ cuprate on both sides of the half-filling 
on the basis of the three-dimensional three-band Hubbard model with a layered structure using the Hartree-Fock approximation. 
Once a small amount of holes or electrons are doped into the half-filled state, our model 
exhibits the charge-transfer insulator-to-metal transition along with a chemical potential jump. 
At the same time, the doped holes or electrons are inhomogeneously distributed, and they tend to form clusters in the vicinity of the half-filling. 
This suggests the possibility of microscopic phase separation with the separation between the metallic and the insulating regions.}

\begin{document}
\sloppy
\maketitle

%
% Introduction
%
\section{Introduction}
As shown by the photoemission spectroscopy, the undoped high-T$_{\mathrm{c}}$ superconducting cuprate (HTSC) is a charge-transfer insulator, 
where the Cu 3d electrons are almost localized by the strong electron correlation.~\cite{Fujimori1987} 
When the rare-earth element is substituted out of the two-dimensional (2D) CuO$_2$ layers and the number of holes or electrons doped into the CuO$_2$ layers is increased, 
Cu 3d electrons hybridized with O 2p electrons achieve itinerancy and display superconductivity. Moreover, in slightly hole-doped La$_{2-x}$Sr$_{x}$CuO$_4$ 
with $0 < x < 0.12$, the chemical potential shift suppression is observed by photoemission spectroscopy (PES).~\cite{Ino1997,Fujimori1998} 
The electronic phase separation between the antiferromagnetic insulating phase and the superconducting phase 
is considered to be one of the reason for the suppression of the shifting of the chemical potential.~\cite{Fujimori2001} 
The phase separation assumes that the doped carriers are inhomogeneously distributed due to the strong electron correlation.

Many experimental findings have suggested that the electrons under such circumstances favor some types of spontaneous ordering in certain doped regions. 
For instance, in order to elucidate the anomalous suppression of the superconducting transition temperature in La$_{1.875}$Ba$_{0.125}$CuO$_4$,~\cite{Moodenbaugh1988} 
the spin and charge correlations in La$_{1.875}$Ba$_{0.125}$CuO$_4$ or (La,Nd)$_{2-x}$Sr$_x$CuO$_4$ with $x\approx0.125$ have been intensively studied, 
and the stripe order has been observed by neutron scattering.~\cite{Tranquada1995,Tranquada1999,Matsuda2002,Christensen2007,Dunsiger2008,Kofu2009,Hucker2011} 
The stripe order has also been observed by x-ray scattering.~\cite{Abbamonte2005,Hucker2010,Hucker2011,Wilkins2011,Dean2013,Fabbris2013} 
Furthermore, an electron paramagnetic resonance study showed that microscopic electronic phase separation occurs in 
La$_{2-x}$Sr$_{x}$Cu$_{0.98}$Mn$_{0.02}$O$_4$ with $0.01 \le x \le 0.06$,~\cite{Shengelaya2004} 
and a Cu nuclear magnetic resonance (NMR) study suggested that a large charge droplet ('blob') 
is formed in the electron-doped Nd$_{1.85}$Ce$_{0.15}$CuO$_{4-\delta}$.~\cite{Bakharev2004} 

Much theoretical works has also been performed to study the behavior of the doped carriers in HTSC. Pioneering works adopting 
the Hartree-Fock approximation (HFA) have studied the stripe order in 
La$_{1.875}$Ba$_{0.125}$CuO$_4$ on the basis of the 2D one-band Hubbard model~\cite{Poilblanc1989,Kato1990} 
or the 2D two-band Hubbard model.~\cite{Zaanen1989} Furthermore, dynamical mean field theory (DMFT) has been exploited 
to study the stripe phase on the basis of the 2D Hubbard model with $L$ non-equivalent sites, where $L=8,\ldots,160$.~\cite{Fleck2001} The DMFT approach 
has also been adopted to analyze the three-band Hubbard model.~\cite{Zolfl2000,Weber2008,Medici2009,Weber2010} In some of these works,~\cite{Weber2008,Weber2010} 
the DMFT approach was combined with the local density approximation (LDA). Another study~\cite{Medici2009} considered the possibility of two-sublattice antiferromagnetism. 
All of these works have successfully reproduced the Zhang-Rice singlet band.~\cite{Zolfl2000,Weber2008,Medici2009,Weber2010} 
This shows that the three-band Hubbard model is an appropriate model for HTSC near the half-filling and that the DMFT is a powerful tool for analyzing its electronic state. 
However, when we investigate the inhomogeneous electronic distribution near the half-filling, we need to adopt the model for a large number of non-equivalent sites. 
In general, the DMFT costs much more than the HFA to analyze the model with a large number of non-equivalent sites.

In this paper, we analyze the normal ground state of the 3D three-band Hubbard model with a single-layered perovskite structure 
in order to study the evolution of the electronic state when holes or electrons are doped into the undoped HTSC. 
We consider 256 non-equivalent copper sites for each rectangular parallelepiped super cell, and adopt the HFA for these conditions. 
We performed the calculation without any assumptions about the electronic distribution, and we obtained fully self-consistent solutions except near the 1/8-filling. 
These solutions showed the chemical potential jump at half-filling, which means that the electron suddenly becomes itinerant when 
a small number of holes or electrons are doped. Moreover, the doped holes or electrons tend to form clusters in the vicinity of the half-filling. 
These clusters are considered to form a metallic region, and are surrounded by the insulating region. 
This suggests the possibility of microscopic electronic phase separation in HTSC near half-fillig.

%
% Formulation
%
\section{Formulation}
\begin{figure}
\includegraphics[width=8.6cm]{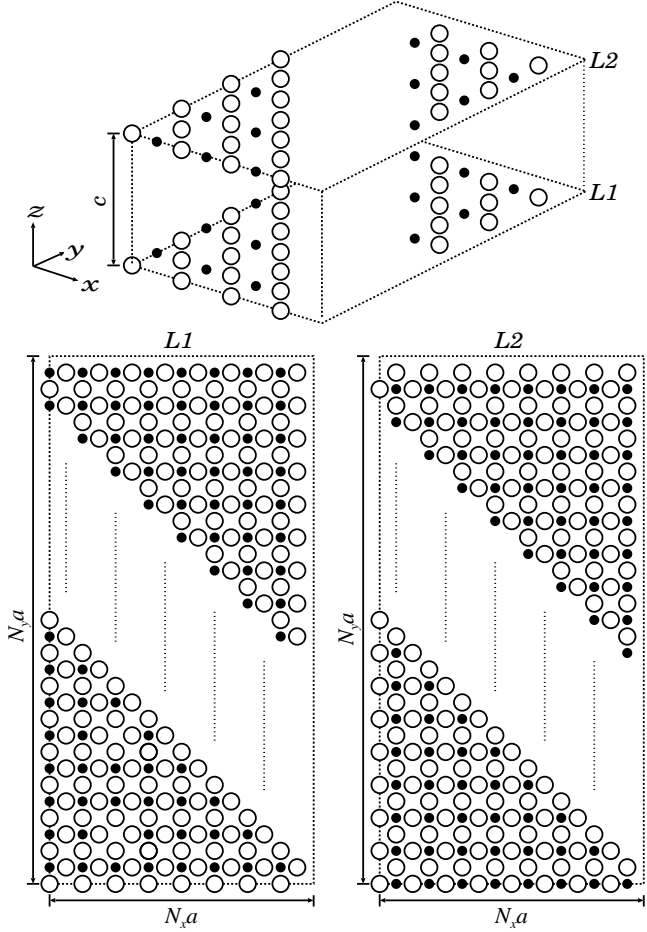}
\caption{\label{figure:1}The schematic figure of the unit cell. $L1$ and $L2$ indicate the two non-equivalent CuO$_2$ layers alternatively stacked along $z$-axis. 
On each CuO$_2$ layer, the size of the unit cell along $x$-axis and $y$-axis is $N_x$ and $N_y$ Cu sites, respectively. In total, the unit cell contains $N_{\mathrm c}$ Cu and $2N_{\mathrm c}$ O sites, where $N_{\mathrm c}=2\times N_x\times N_y$.}
\end {figure}

Our 3D three-band Hubbard model Hamiltonian, $\hat{H}$, is composed of {\it d}-electrons at each Cu site and 
{\it p}-electrons at each O site. To consider the spatial inhomogeneity, we introduce the rectangular parallelepiped super cell containing 
$N_{\mathrm c}$ Cu and $2N_{\mathrm c}$ O sites as a unit cell. Thus, $\hat{H}$ is defined as follows:
\begin{equation}
\hat{H} = \sum_{i=1}^{N_{\mathrm c}}\sum_{j=1}^{N_{\mathrm c}}\sum_{{\mathbf k}\sigma}\hat{C}^\dagger_{i{\mathbf k}\sigma}\hat{H}_{ij {\mathbf k}}\hat{C}_{j{\mathbf k}\sigma} + \frac{U}{N} \sum_{i=1}^{N_{\mathrm c}}\sum_{{\mathbf k}  {\mathbf k}^\prime {\mathbf q}}
d_{i {\mathbf k}+{\mathbf q} \uparrow}^\dagger d_{i {\mathbf k}^\prime-{\mathbf q} \downarrow}^\dagger d_{i {\mathbf k}^\prime \downarrow} d_{i {\mathbf k} \uparrow} 
- \mu \sum_{i=1}^{N_{\mathrm c}}\sum_{{\mathbf k}\sigma}\hat{C}^\dagger_{i {\mathbf k} \sigma}\hat{C}_{i {\mathbf k} \sigma}.
\label{eq:1}
\end{equation}
Here we use the abbreviations $\hat{C}^\dagger_{i{\mathbf k}\sigma} \equiv (d^\dagger_{i{\mathbf k}\sigma}\, p^{x\dagger}_{i{\mathbf k}\sigma}\, p^{y\dagger}_{i{\mathbf k}\sigma})$ 
and $\hat{C}_{i{\mathbf k}\sigma} \equiv \,^t\!(d_{i{\mathbf  k}\sigma}\, p^x_{i{\mathbf  k}\sigma}\, p^y_{i{\mathbf  k}\sigma})$, where 
$d_{i {\mathbf k} \sigma}(d_{i {\mathbf k} \sigma}^\dagger)$ and $p_{i {\mathbf k} \sigma}^{x(y)}(p_{i{\mathbf k} \sigma}^{x(y) \dagger})$ are 
the annihilation (creation) operators for the $d$-orbital and $p^{x(y)}$-orbital electron on the $i$-th site, 
as specified by the momentum ${\mathbf k}$ and spin $\sigma=\{\uparrow,\downarrow\}$, respectively. 
$U$, $N$, and $\mu$ are the on-site Coulomb repulsion between the $d$-orbitals, 
the number of ${\mathbf k}$-space lattice points in the first Brillouin zone (FBZ), and the chemical potential, respectively. 
The FBZ is defined in the reciprocal space to the lattice whose unit cell contains $N_{\mathrm c}$ Cu and $2N_{\mathrm c}$ O sites. 
The unit cell is schematically shown in Fig.~\ref{figure:1}. The two non-equivalent CuO$_2$ layers, indicated with $L1$ and $L2$, 
are alternatively stacked along $z$-axis. On each CuO$_2$ layer, the size of the unit cell along $x$-axis and $y$-axis is $N_x$ and $N_y$ Cu sites, respectively. 
Thus, when we set $N_{\mathrm x}=8$ and $N_{\mathrm y}=16$, $N_{\mathrm c}=2 \times 8 \times 16=256$. 
We take the primitive translation vectors for the unit cell to be $\pm 8a(\hat{\mathbf x} + \hat{\mathbf y})$, $\pm 8a(\hat{\mathbf x} - \hat{\mathbf y})$, and $\pm 2c\hat{\mathbf z}$, 
where $\hat{\mathbf x}$, $\hat{\mathbf y}$, and $\hat{\mathbf z}$ are the unit vectors for the $x$-axis, $y$-axis, and $z$-axis, respectively. 
$\hat{H}_{ij {\mathbf k}}$ in Eq.~(\ref{eq:1}) is defined as follows:
\begin{equation}
\hat{H}_{ij {\mathbf k}} =
\left(
\begin{array}{ccc}
\Delta_{dp}\delta_{\mathbf{R}_{i d}\mathbf{R}_{j d}} + \zeta_{ij {\mathbf k}}^z & \zeta_{ij {\mathbf k}}^x  & \zeta_{ij {\mathbf k}}^y \\ 
\zeta_{ji {\mathbf k}}^{x*} & 0 & \zeta_{ij {\mathbf k}}^p \\
\zeta_{ji {\mathbf k}}^{y*} & \zeta_{ji {\mathbf k}}^{p*} & 0 \\
\end{array} \right),
\end{equation}
where $\Delta_{dp}$ is the hybridization gap energy between the $d$- and $p^{x(y)}$-orbitals. $\mathbf{R}_{i\varphi}$, 
where $\varphi \in (d,\,p^x,\,p^y)$, is the coordinate of the $\varphi$-orbital electron on the $i$-th site. $\delta_{\mathbf{R}_{i\varphi}\mathbf{R}_{j\varphi^\prime}}$ is Kronecker's delta, 
i.e., it is $1$ for $\mathbf{R}_{i\varphi}=\mathbf{R}_{j\varphi^\prime}$ and $0$ for $\mathbf{R}_{i\varphi}\neq\mathbf{R}_{j\varphi^\prime}$. 
In the following, we take both $a$ and $c$ to be a unit of length and set $a=c=1$. Then, we can represent
\begin{eqnarray*}
\zeta_{ij {\mathbf k}}^p & = & t_{pp}\left[ e^{i(k_x/2-k_y/2)}\delta_{\mathbf{R}_{i p^x}\mathbf{R}_{j p^y}+\hat{\mathbf x}/2-\hat{\mathbf y}/2}
+ e^{-i(k_x/2-k_y/2)}\delta_{\mathbf{R}_{i p^x}\mathbf{R}_{j p^y}-\hat{\mathbf x}/2+\hat{\mathbf y}/2} \right. \\
& & \left.\hspace{2em}- e^{i(k_x/2+k_y/2)}\delta_{\mathbf{R}_{i p^x}\mathbf{R}_{j p^y}+\hat{\mathbf x}/2+\hat{\mathbf y}/2}
- e^{-i(k_x/2+k_y/2)}\delta_{\mathbf{R}_{i p^x}\mathbf{R}_{j p^y}-\hat{\mathbf x}/2-\hat{\mathbf y}/2} \right],
\end{eqnarray*}
\[\zeta_{ij {\mathbf k}}^x=t_{dp}\left[e^{ik_x/2}\delta_{\mathbf{R}_{i d}\mathbf{R}_{j p^x}+\hat{\mathbf x}/2} - e^{-ik_x/2}\delta_{\mathbf{R}_{i d}\mathbf{R}_{j p^x}-\hat{\mathbf x}/2}\right],\]
\[\zeta_{ij {\mathbf k}}^y=t_{dp}\left[e^{ik_y/2}\delta_{\mathbf{R}_{i d}\mathbf{R}_{j p^y}+\hat{\mathbf y}/2} - e^{-ik_y/2}\delta_{\mathbf{R}_{i d}\mathbf{R}_{j p^y}-\hat{\mathbf y}/2}\right],\] and
\begin{eqnarray*}
\zeta_{ij {\mathbf k}}^z & = & t_\perp\left[ e^{i(k_x/2-k_y/2+k_z)}\delta_{\mathbf{R}_{i d}\mathbf{R}_{j d}+\hat{\mathbf x}/2-\hat{\mathbf y}/2+\hat{\mathbf z}}
+ e^{-i(k_x/2-k_y/2-k_z)}\delta_{\mathbf{R}_{i d}\mathbf{R}_{j d}-\hat{\mathbf x}/2+\hat{\mathbf y}/2+\hat{\mathbf z}} \right. \\
& & \hspace{2em}+ e^{i(k_x/2+k_y/2+k_z)}\delta_{\mathbf{R}_{i d}\mathbf{R}_{j d}+\hat{\mathbf x}/2+\hat{\mathbf y}/2+\hat{\mathbf z}} 
+ e^{-i(k_x/2+k_y/2-k_z)}\delta_{\mathbf{R}_{i d}\mathbf{R}_{j d}-\hat{\mathbf x}/2-\hat{\mathbf y}/2+\hat{\mathbf z}} \\
& & \hspace{2em}+ e^{i(k_x/2-k_y/2-k_z)}\delta_{\mathbf{R}_{i d}\mathbf{R}_{j d}+\hat{\mathbf x}/2-\hat{\mathbf y}/2-\hat{\mathbf z}} 
+ e^{-i(k_x/2-k_y/2+k_z)}\delta_{\mathbf{R}_{i d}\mathbf{R}_{j d}-\hat{\mathbf x}/2+\hat{\mathbf y}/2-\hat{\mathbf z}} \\
& & \left.\hspace{2em}+ e^{i(k_x/2+k_y/2-k_z)}\delta_{\mathbf{R}_{i d}\mathbf{R}_{j d}+\hat{\mathbf x}/2+\hat{\mathbf y}/2-\hat{\mathbf z}}
+ e^{-i(k_x/2+k_y/2+k_z)}\delta_{\mathbf{R}_{i d}\mathbf{R}_{j d}-\hat{\mathbf x}/2-\hat{\mathbf y}/2-\hat{\mathbf z}} \right],
\end{eqnarray*} 
where $t_{pp}$ is the transfer energy between a
{\it p}$^x$-orbital and a {\it p}$^y$-orbital, $t_{dp}$ is that between a {\it d}-orbital and a {\it p}$^{x(y)}$-orbital, 
and $t_\perp$ is that between {\it d}-orbitals, respectively. In this study, $t_{dp}$ is the unit of energy.

We adopt the HFA with respect to every $N_{\mathrm c}$ Cu site and two spin states, 
and we only consider collinear spin states. Thus, we define
\begin{equation}
\frac{1}{N}\left<d_{i{\mathbf k} \sigma}^\dagger d_{i{\mathbf k}^\prime \sigma^\prime} \right>
\equiv n_{i d \sigma} \delta_{{\mathbf k}{\mathbf k}^\prime}\delta_{\sigma \sigma^\prime},
\label{eq:2}
\end{equation}
and we approximate Eq.~(\ref{eq:1}) as follows:
\begin{equation}
\hat{H} \approx \sum_{i=1}^{N_{\mathrm c}}\sum_{j=1}^{N_{\mathrm c}}\sum_{{\mathbf k}\sigma}\hat{C}^\dagger_{i{\mathbf k}\sigma}\hat{H}_{ij {\mathbf k}\sigma}^{\mathrm{U}}\hat{C}_{j{\mathbf k}\sigma} 
- \mu \sum_{i=1}^{N_{\mathrm c}}\sum_{{\mathbf k}\sigma}\hat{C}^\dagger_{i {\mathbf k} \sigma}\hat{C}_{i {\mathbf k} \sigma},
\label{eq:3}
\end{equation}
\begin{equation}
\hat{H}_{ij {\mathbf k}\sigma}^{\mathrm{U}} =
\left(
\begin{array}{ccc}
\left[\Delta_{dp}+Un_{i d -\sigma}\right]\delta_{\mathbf{R}_{i d}\mathbf{R}_{j d}} + \zeta_{ij {\mathbf k}}^z & \zeta_{ij {\mathbf k}}^x  & \zeta_{ij {\mathbf k}}^y \\ 
\zeta_{ji {\mathbf k}}^{x*} & 0 & \zeta_{ij {\mathbf k}}^p \\
\zeta_{ji {\mathbf k}}^{y*} & \zeta_{ji {\mathbf k}}^{p*} & 0 \\
\end{array} \right).
\label{eq:4}
\end{equation}
Then, we conduct a self-consistent calculation of Eqs.~(\ref{eq:2}), (\ref{eq:3}), and (\ref{eq:4}) and obtain self-consistent fields $n_{i d \sigma}$, $n_{i p^x \sigma}$, and 
$n_{i p^y \sigma}$, where we define
\begin{equation}
\frac{1}{N}\left<p_{i{\mathbf k} \sigma}^{x\dagger} p_{i{\mathbf k}^\prime \sigma^\prime}^x \right>
\equiv n_{i p^x \sigma} \delta_{{\mathbf k}{\mathbf k}^\prime}\delta_{\sigma \sigma^\prime},
\end{equation}
\begin{equation}
\frac{1}{N}\left<p_{i{\mathbf k} \sigma}^{y\dagger} p_{i{\mathbf k}^\prime \sigma^\prime}^y \right>
\equiv n_{i p^y \sigma} \delta_{{\mathbf k}{\mathbf k}^\prime}\delta_{\sigma \sigma^\prime}.
\end{equation}
These satisfy 
\begin{equation}
\sum_{i=1}^{N_{\mathrm c}}\left[5-\sum_{\sigma}(n_{i d \sigma}+n_{i p^x \sigma}+n_{i p^y \sigma})\right] = N_{\mathrm c}\delta_{\mathrm h}
\end{equation}
for a given total number of doped holes $\delta_{\mathrm h}$.

%
%Results and Discussion
%
\section{Results and Discussion}
In the numerical calculations, we divide the FBZ into $N=16 \times 32 \times 4$ equally-spaced rectangular parallelepiped. 
The parameter sets are selected as $t_{dp}=1.0\,{\mathrm{eV}}$, $t_{pp}=-0.3\,{\mathrm{eV}}$, $t_\perp=0.005\,{\mathrm{eV}}$, $\Delta_{dp}=0.0\,{\mathrm{eV}}$, and 
$U=6.0\,{\mathrm{eV}}$. In order to obtain self-consistent solutions, we need to set initial values for fields $n_{i d \sigma}$, $n_{i p^x \sigma}$, and $n_{i p^y \sigma}$ and carry out 
iterative calculation until all of these fields have sufficient accuracy. Here, we chose the initial values for the fields from the uniform-distributed random numbers. 
In this manner, we obtained the solutions for the doping region near half-filling; 
$0.0 \leq \delta_{\mathrm{h}} \leq 0.1$ and $0.0 \leq \delta_{\mathrm{e}} \leq 0.1$, where $\delta_{\mathrm{e}} \equiv -\delta_{\mathrm{h}}$. 
For these solutions, all of the obtained fields $n_{i d \sigma}$, $n_{i p^x \sigma}$, and $n_{i p^y \sigma}$ have three digits of accuracy. 
Once they get three digits of accuracy, all of these fields rapidly converge. Thus, we can consider the solutions with these fields as fully self-consistent. 
We also tried to obtain the solutions for other doping region e.g. $\delta_{\mathrm{h}} \sim 0.125$ but failed. 
For these doping regions, the ground state accompanied by certain types of long-period superlattice structure would be stable. 
The fields corresponding to such ground state could be hardly obtained by our iterative calculation. We henceforth concentrate our discussion on the doping region near half-filling. 

\begin{figure}
\includegraphics[width=8.6cm]{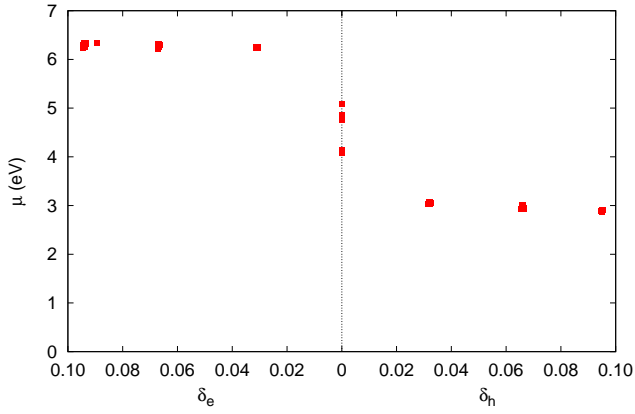}
\caption{\label{figure:2}(Color online) The doping dependence of the chemical potential $\mu$. $\delta_{\mathrm{e}}$ and $\delta_{\mathrm{h}}$ are the numbers of doped electrons and holes per CuO$_2$ unit, respectively.}
\end {figure}

\begin{figure}
\includegraphics[width=8.6cm]{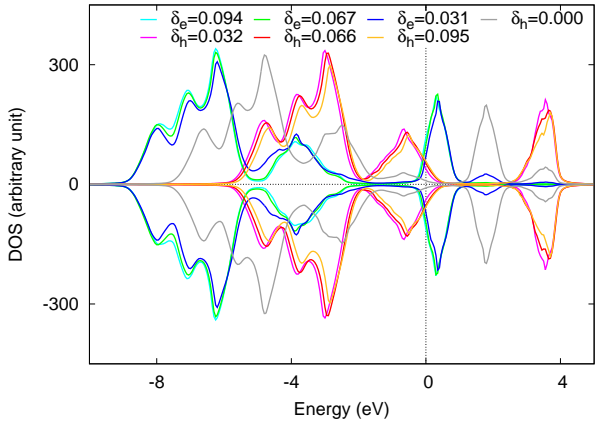}
\caption{\label{figure:3}(Color) The doping dependence of the mean DOS. The Fermi level at zero energy is indicated by the vertical dashed line. The lines of upper half and lower half indicate the DOS for spin up and spin down, respectively. In the ascending orders of energy, we attribute each three blocks of DOS to {\it p}-band plus lower Hubbard {\it d}-band, {\it p}-band, and upper Hubbard {\it d}-band, respectively.}
\end {figure}

The doping dependence of the chemical potential $\mu$ for our fully self-consistent solutions is shown in Fig.~\ref{figure:2}. 
The chemical potential changes rapidly at half-filling, which is consistent with experimental results obtained by comprehensive PES studies on HTSC.~\cite{Fujimori2002} 
The magnitude of this change is about $3.0\,{\mathrm{eV}}$, and the value is in accordance with $U/2-\Delta_{dp}$, which is equal to $U/2$ in our calculations. 
This fact can be explained by the doping dependence of the density of states (DOS), which is shown in Fig.~\ref{figure:3}. 
In this figure, we show the mean DOS over the solutions with similar doping and chemical potential as DOS for each doping state, 
and we indicate each doping state with the mean doping over these solutions. All these labels in Fig.~\ref{figure:3} are summarized in Table~\ref{table:1} 
For instance, DOS for $\delta_{\mathrm{e}}=0.094$ in Fig.~\ref{figure:3} is 
the mean DOS over the seven solutions for which we found $\delta_{\mathrm{e}}=0.094\pm0.000$ and $\mu=6.301\pm0.015$ 
as a mean with standard error of the mean of doping and chemical potential, respectively.
\begin{table*}
\begin{tabular}{rccc}
\hline
 & N & $\delta_{\mathrm{(e,h)}} \pm {\mathrm{SEM}}$ & $\mu \pm {\mathrm{SEM}}$ \\ 
\hline
$\delta_{\mathrm{e}}=0.094$ & $7$ & $0.094 \pm 0.000$ & $6.301 \pm 0.015$ \\
$\delta_{\mathrm{e}}=0.067$ & $8$ & $0.067 \pm 0.000$ & $6.287 \pm 0.011$ \\
$\delta_{\mathrm{e}}=0.031$ & $8$ & $0.031 \pm 0.000$ & $6.246 \pm 0.006$ \\
$\delta_{\mathrm{h}}=0.000$ & $5$ & $0.000 \pm 0.000$ & $4.823 \pm 0.019$ \\
$\delta_{\mathrm{h}}=0.032$ & $8$ & $0.032 \pm 0.000$ & $3.054 \pm 0.005$ \\
$\delta_{\mathrm{h}}=0.066$ & $8$ & $0.066 \pm 0.000$ & $2.971 \pm 0.011$ \\
$\delta_{\mathrm{h}}=0.095$ & $8$ & $0.095 \pm 0.000$ & $2.892 \pm 0.004$ \\
\hline
\end{tabular}
\caption{\label{table:1}The statistical values of the solutions with similar doping $\delta_{\mathrm{(e,h)}}$ and chemical potential $\mu$, which are selected from the ones in Fig.~\ref{figure:2}. N and SEM mean number of solutions and standard error of the mean, respectively. These labels are commonly used in Figs.~\ref{figure:3}, \ref{figure:4}, \ref{figure:5}, \ref{figure:6}, and \ref{figure:7}.}
\end{table*}
In the slightly electron-doped case, $\delta_{\mathrm{e}}=0.094, 0.067, 0.031$, the Fermi level crosses the upper Hubbard {\it d}-band, 
while in the slightly hole-doped case, $\delta_{\mathrm{h}}=0.032, 0.066, 0.095$, 
the Fermi level crosses the {\it p}-band. The upper {\it d}-level is about $Un_{i d \sigma}/2$ larger than the {\it p}-level 
when $\Delta_{dp}=0.0\,{\mathrm{eV}}$, due to the Hubbard splitting. In the energy range between the {\it p}-band and the upper Hubbard {\it d}-band, 
the density of states is almost zero. In the undoped case, $\delta_{\mathrm{h}}=0.000$, the Fermi level locates in this energy gap, and it causes the chemical potential to jump by $U/2$ at half-filling. 
The quantum Monte-Carlo calculation of the two-band Hubbard model in the limit of infinite dimensions gave the same result for the chemical potential jump 
as our calculation.~\cite{Georges1993} Thus, the chemical potential jump at half-filling is the characteristic behavior of the multi-band Hubbard model 
composed of both Cu 3d electrons and O 2p electrons independent of the dimensionality of the model.

\begin{figure}
\includegraphics[width=8.6cm]{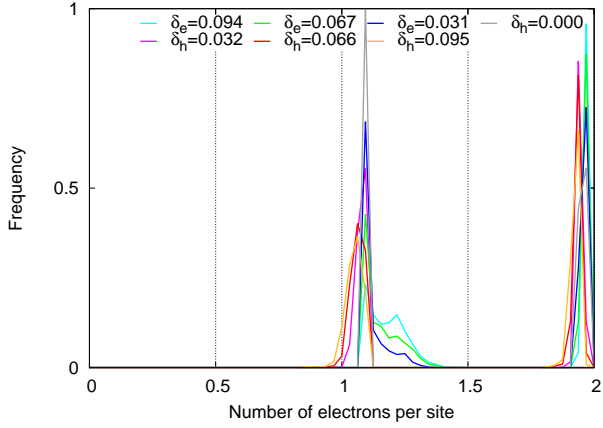}
\caption{\label{figure:4}(Color) The doping dependence of the histogram for the number of electrons per site. We can attribute the peaks whose number of electrons per site are at or near 1 to the ones for $n_{i d \sigma}$ and those whose number of electrons per site are at or near 2 to the ones for $n_{i p^x \sigma}$ or $n_{i p^y \sigma}$.}
\end {figure}

The inhomogeneous distribution of every obtained field $n_{i \varphi \sigma}$ can be observed in our solutions. 
The doping dependence of the distribution of the obtained fields can be shown by histogram, as in Fig.~\ref{figure:4}. 
In this figure, we show the accumulated histograms over the solutions with similar doping and chemical potential as the histogram for each doping state, 
and we indicate each doping state with the mean doping over these solutions as well as in Fig.~\ref{figure:3}. 
At half-filling, where the ground state is insulating, almost all $n_{i d \sigma}$ have the same value near $1$ 
and almost all $n_{i p^x \sigma}$ and $n_{i p^y \sigma}$ have the same value near $2$. 
In the electron-doped case, with the increase of the doped electrons, only the peak for $n_{i d \sigma}$ broadens and its center shifts higher. 
In contrast, in the hole-doped case, with increasing hole density, not only the peak for $n_{i d \sigma}$ but also the peaks for 
$n_{i p^x \sigma}$ and $n_{i p^y \sigma}$ broaden and their centers shift lower. Except for their variances, the doping dependence of the average of $n_{i d \sigma}$ is similar to 
that of $n_d$ obtained by the LDA+DMFT calculation, and the doping dependence of the average of $n_{i p^x \sigma}$ or $n_{i p^y \sigma}$ is similar to that of $n_p$.~\cite{Weber2008}
\begin{figure}
\includegraphics[width=8.6cm]{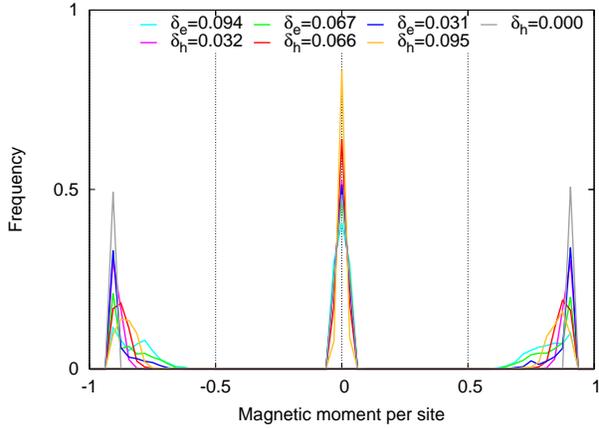}
\caption{\label{figure:5}(Color) The doping dependence of the histogram for the magnetic moment per site. We can attribute the peaks whose magnetic moment per site are at or near $\pm1$ to the ones for $m_{i d}$ and those whose magnetic moment per site are at or near 0 to the ones for $m_{i p^x}$ or $m_{i p^y}$.}
\end {figure}

The spacial distribution of the magnetic moment $m_{i \varphi} \equiv n_{i \varphi \uparrow} - n_{i \varphi \downarrow}$ also becomes inhomogeneous by doping. 
In the same manner as in Fig.~\ref{figure:4}, the doping dependence of the distribution of the magnetic momenta magnitude can be shown by histogram in Fig.~\ref{figure:5}.  
At half-filling, one half of $m_{i d}$ is exactly at $1$ and the other half is exactly at $-1$. This indicates that the {\it d}-electron spins are fully polarized. 
Both in the electron-doped case and in the hole-doped case, with an increase of the doped carrier, the two peaks for $m_{i d}$ become broader and their centers shift toward zero. 
This means that the number of the fully-polarized {\it d}-electron spins decreases and that the number of the doubly-occupied Cu sites increases. We should note that the tails of $m_{i d}$ 
in the electron-doped case extend more than those in the hole-doped case.

\begin{figure}
\includegraphics[width=8.6cm]{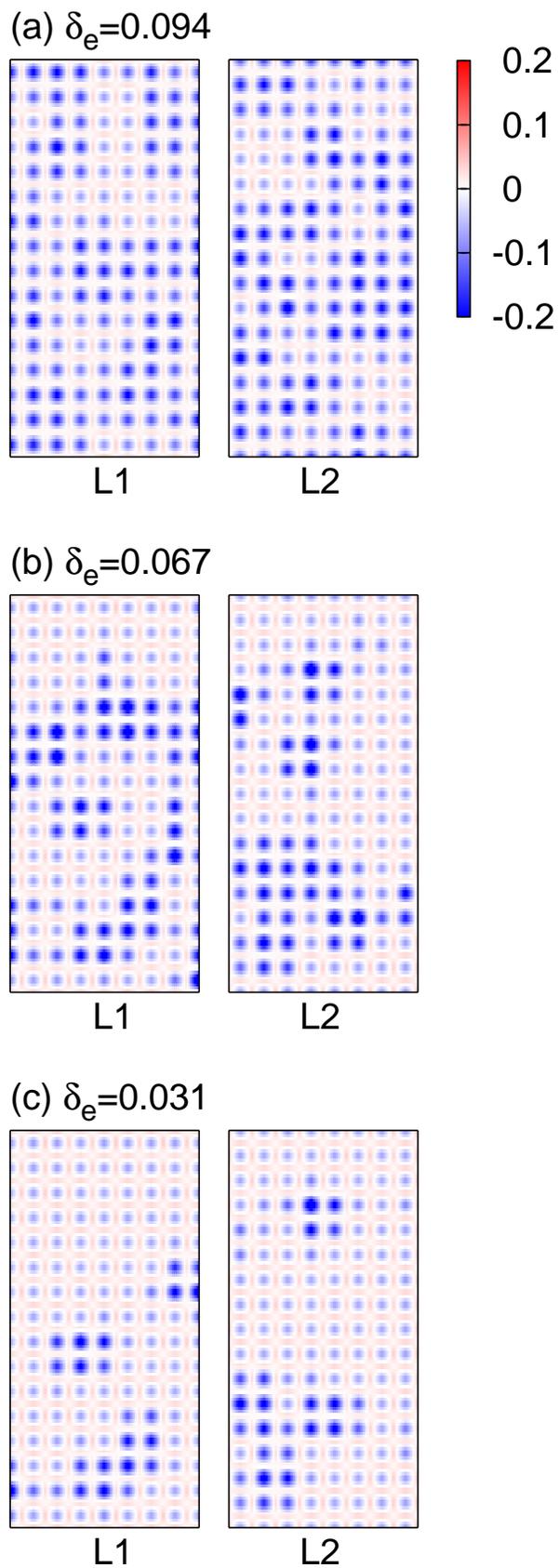}
\caption{\label{figure:6}(Color online) The doping dependence of $\delta_{\mathrm{h}}(\mathbf{r})$ for the electron-doped case. Each figure is a single snapshot of the solutions attributed to each doping state, indicated with the mean doping over these solutions.}
\end {figure}

\begin{figure}
\includegraphics[width=8.6cm]{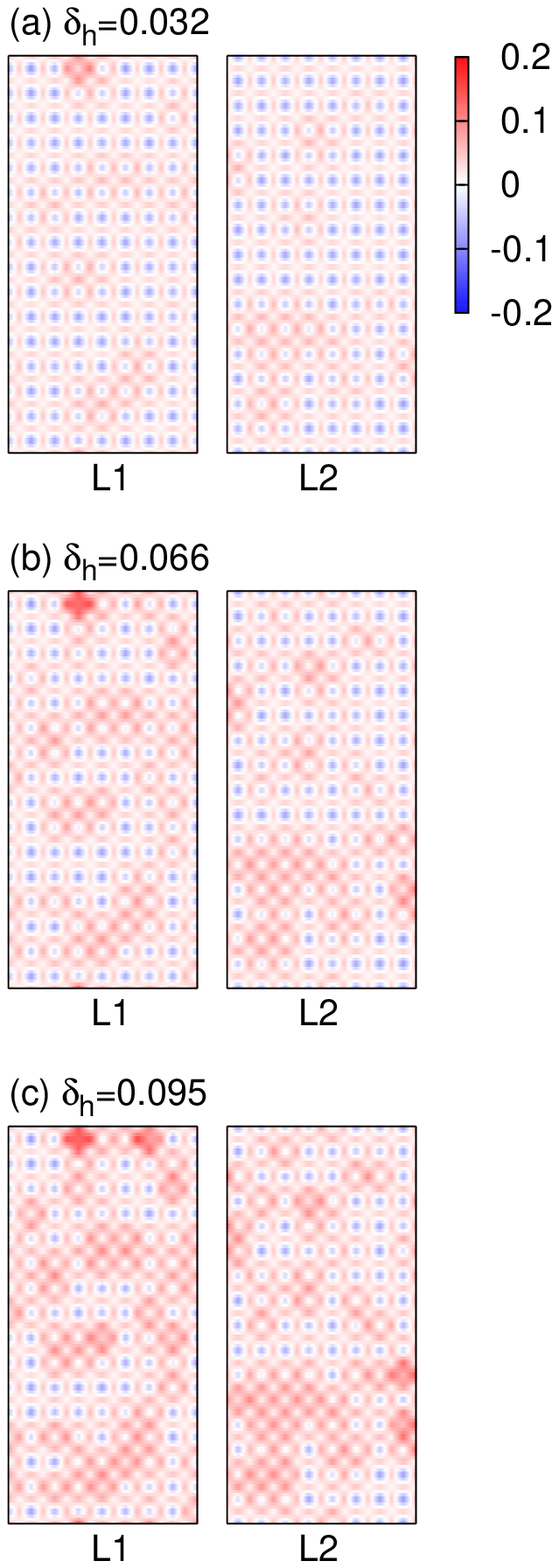}
\caption{\label{figure:7}(Color online) The doping dependence of $\delta_{\mathrm{h}}(\mathbf{r})$ for the hole-doped case. Each figure is a single snapshot of the solutions attributed to each doping state, indicated with the mean doping over these solutions.}
\end {figure}

\begin{figure}
\includegraphics[width=8.6cm]{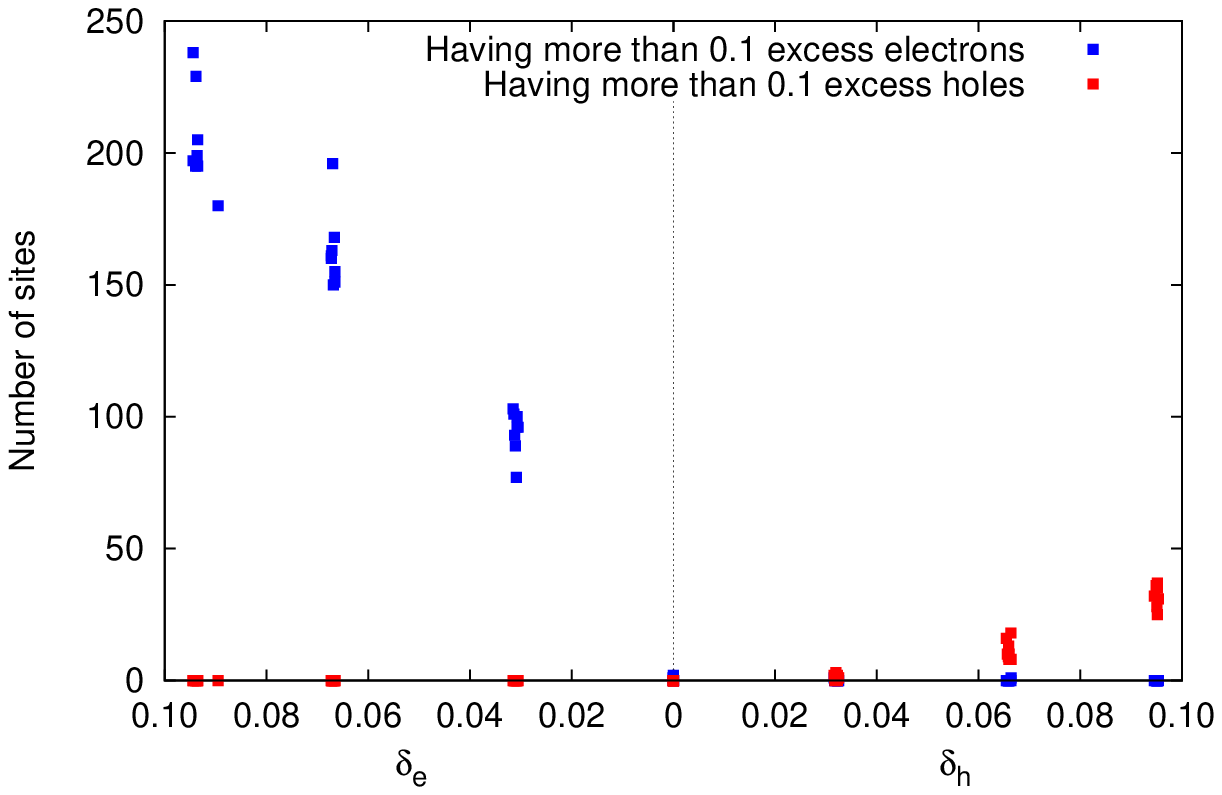}
\caption{\label{figure:8}(Color online) The doping dependence of the number of the sites having more than 0.1 excess carriers.}
\end {figure}

The difference between the electron-doped case and the hole-doped case is caused by the doped carriers being differently distributed in the unit cell. 
In order to show this distribution, we define the distribution function of the doped carriers as follows:
\begin{equation}
\delta_{\mathrm{h}}(\mathbf{r}) \equiv \sum_i\left[(1-n_{i d \uparrow}-n_{i d \downarrow})\delta_{\mathbf{r}\mathbf{R}_{i d}}\hspace{1em}+\hspace{-1em}\sum_{\varphi \in (p^x,\,p^y)}\hspace{-1em}(2-n_{i \varphi \uparrow}-n_{i \varphi \downarrow})\delta_{\mathbf{r}\mathbf{R}_{i \varphi}}\right].
\end{equation}
The doping dependences of $\delta_{\mathrm{h}}(\mathbf{r})$ for the electron-doped case and for the hole-doped cases are shown 
in Figs.~\ref{figure:6} and~\ref{figure:7}, respectively. 
In these figures, the red spots indicate the location of doped {\it holes} and the blue ones indicate the location of doped {\it electrons}. 
In the electron-doped case, as shown in Fig.~\ref{figure:6}, the doped electrons form blobs even in the slightly electron-doped case, with 
$\delta_{\mathrm{e}}=0.031$. These blobs can be identified as the ones in Nd$_{1.85}$Ce$_{0.15}$CuO$_{4-\delta}$, suggested by the Cu NMR study.~\cite{Bakharev2004} 
On the other hand, as shown in Fig.~\ref{figure:7}, the doped holes stay within a single CuO$_4$ cluster in the slightly hole-doped case, with $\delta_{\mathrm{h}}=0.032$. 
Hence it can be recognized that the cluster formed by the doped electrons is larger than the one formed by the doped holes. We can quantitatively clarify this tendency 
with the doping dependence of the number of the sites having more than certain value of excess carriers in Fig.~\ref{figure:8} 
since such a number should increase with the size of the cluster formed by the doped carriers. 
The difference between the electron-doped case and the hole-doped case is explained in the following; In both cases, the doped carriers basically tend to form an extended cluster, since the strong on-site Coulomb repulsion hinders the double occupancy on each site, and instead align them next to each other to gain more kinetic energy. The study on Coulomb gas ordering in a 3D layered system by the Brownian dynamics approach supports our explanation.~\cite{Pashkevich2001} In our lattice model, the size of such a cluster depends on the number of adjacent orbitals where the doped carriers are allowed to occupy. In the electron-doped case, 
the doped electrons have room to sit only on the Cu sites, because all O sites are fully filled by electrons. On the other hand, in the hole-doped case, the doped holes have room to sit on both Cu sites and O sites. Therefore, when the number of doped electrons is as many as the number of doped holes, the cluster formed by the doped electrons is larger than the one formed by the doped holes.

%
% Conclusion
%
\section{Conclusion}
In this paper, we conducted HFA calculations for the 3D three-band Hubbard model 
with a single-layered perovskite structure considering a large number of non-equivalent sites.
We obtained the fully self-consistent solutions both for the electron-doped and hole-doped cases at or near half-filling. 
Our solutions show the chemical potential jump at half-filling. The jump can be explained by the DOS dependence on doping, 
which is characteristic to the multi-band Hubbard model composed of both Cu 3d electrons and O 2p electrons 
independent of the dimensionality of the model. Inhomogeneous electronic distributions near half-filling are observed in our solutions. 
There is a remarkable difference in the inhomogeneous electronic distributions between the electron-doped and hole-doped cases. 
That is, the clusters formed by doped carriers extend more in the electron-doped cases than in the hole-doped cases. 
The difference between the electron-doped and hole-doped case is caused by the difference in the species of orbitals 
the electron and hole are allowed to occupy, which should be explained only on the basis of the multi-band Hubbard model. 
Thus, the theoretical approach on the basis of the multi-band Hubbard model can explain 
both the chemical potential jump at half-filling and inhomogeneous electronic distributions near half-filling in a comprehensive way.

\section*{Acknowledgments}
The authors are grateful to Dr. N. Nakai and Prof. Y. Aiura for their stimulating discussions. We are also grateful to an anonymous reviewer of the first manuscript 
for providing insightful comments and directions which have resulted in the revised manuscript.

\end{document}